\RequirePackage{fix-cm}

\documentclass{svjour3}                     % onecolumn (standard format)
\smartqed  % flush right qed marks, e.g. at end of proof
 \usepackage{mathptmx}      % use Times fonts if available on your TeX system
 \usepackage{aps-bibstyle}  % use this style if you upload to .tex file only a part of Bibtex created bbl.
\usepackage[dvipsnames]{xcolor}
\usepackage[colorlinks=true,urlcolor=Blue,linkcolor=Blue]{hyperref}
\usepackage[all]{hypcap}
\usepackage{latexsym}
\journalname{my journal}
\usepackage{epsfig}
\usepackage{graphicx}

\begin{document}

\title{SPARC Collaboration: New Strategy for Storage Ring Physics at FAIR%\thanks{Grants or other notes
}


\author{%
Thomas St\"ohlker \and
Yuri A. Litvinov \and
Angela Br\"auning-Demian \and
Michael Lestinsky \and
Frank Herfurth \and
Rudolf Maier \and
Dieter Prasuhn \and
Reinhold Schuch \and
%Rolf Stassen \and
Markus Steck \and
%Hans Stockhorst \and
for the SPARC Collaboration
}

\institute{
Thomas St\"ohlker \at Helmholtz-Institut Jena, 07743 Jena, Germany \email{T.Stoehlker@gsi.de} \and
Thomas St\"ohlker \and Yuri A. Litvinov \and Angela Br\"auning-Demian \and Michael Lestinsky \and Frank Herfurth \and Markus Steck \at
GSI Helmholtzzentrum f{\"u}r Schwerionenforschung, 64291 Darmstadt, Germany \and
Thomas St\"ohlker \at Friedrich-Schiller-Universit{\"a}t Jena, 07737 Jena, Germany \and
Reinhold Schuch \at Department of Atomic Physics, Stockholm University, AlbaNova, 10691 Stockholm, Sweden \and
%Rudolf Maier \and Dieter Prasuhn \and Rolf Stassen \and Hans Stockhorst \at Forschungszentrum J{\"u}lich, 52428 J{\"u}lich, Germany
Rudolf Maier \and Dieter Prasuhn \at Forschungszentrum J{\"u}lich, 52428 J{\"u}lich, Germany
}%

\authorrunning{Thomas St\"ohlker et al.}

\date{Received: date / Accepted: date}

\maketitle

\begin{abstract}
SPARC collaboration at FAIR pursues the worldwide unique research program
by utilizing storage ring and trapping facilities for highly-charged heavy ions.
The main focus is laid on the exploration of the physics at strong, ultra-short electromagnetic fields
including the fundamental interactions between electrons and heavy nuclei as well
as on the experiments at the border between nuclear and atomic physics.
Very recently SPARC worked out a realization scheme for experiments with highly-charged heavy-ions at relativistic energies in the
High-Energy Storage Ring HESR and at very low-energies at the CRYRING coupled to the present ESR.
Both facilities
%will become available within the Modularised Start Version of FAIR and will thus
provide unprecedented physics opportunities already at the very early stage of FAIR operation.
The installation of CRYRING, dedicated Low-energy Storage Ring (LSR) for FLAIR,
may even enable a much earlier realisation of the physics program of FLAIR with slow anti-protons.
\end{abstract}

\section{Introduction}
\label{intro}
Highest intensities for relativistic beams of stable and unstable heavy nuclei are expected at the international Facility for Antiproton and Ion Reserach, FAIR project~\cite{FAIR1}.
The heart of FAIR will be the two synchrotrons, SIS-100 and SIS-300~\cite{CDR},
which will be able to accelerate intense beams of protons and heavy ions,
provided by the presently operating at GSI synchrotron SIS-18 \cite{SIS18}, to the highest magnetic rigidities of 100 and 300~Tm, respectively.
The ion beams from the synchrotrons will be available for various experimental programs combined into four FAIR research pillars,
namely the Compressed Barionic Matter experiment (CBM) to explore the QCD phase diagram at high baryon densities,
the anti{P}roton {AN}nihilation at {DA}rmstadt (PANDA) experiment to investigate the structure of hadronic matter,
the Nuclear Structure and Astrophysics (NuSTAR) experiments aiming at studying properties of exotic nuclei,
and the pillar representing  {A}tomic Physics, Biophysics, {P}lasma {P}hysics and {A}pplications (APPA).
APPA  is an umbrella organization, consisting out of the five collaborations
BIOMAT (Biology and Material Science), FLAIR (Facility for Low-Energy Antiproton and Heavy Ion Research),
HEDgeHOB (High Energy Density Matter generated by Heavy Ion Beams), WDM (Warm Dense Matter) and
SPARC (Stored Particles Atomic Research Collaboration) collaborations, with latter being the main focus of the present paper.

\begin{figure}[t!]
\includegraphics[width=\textwidth]{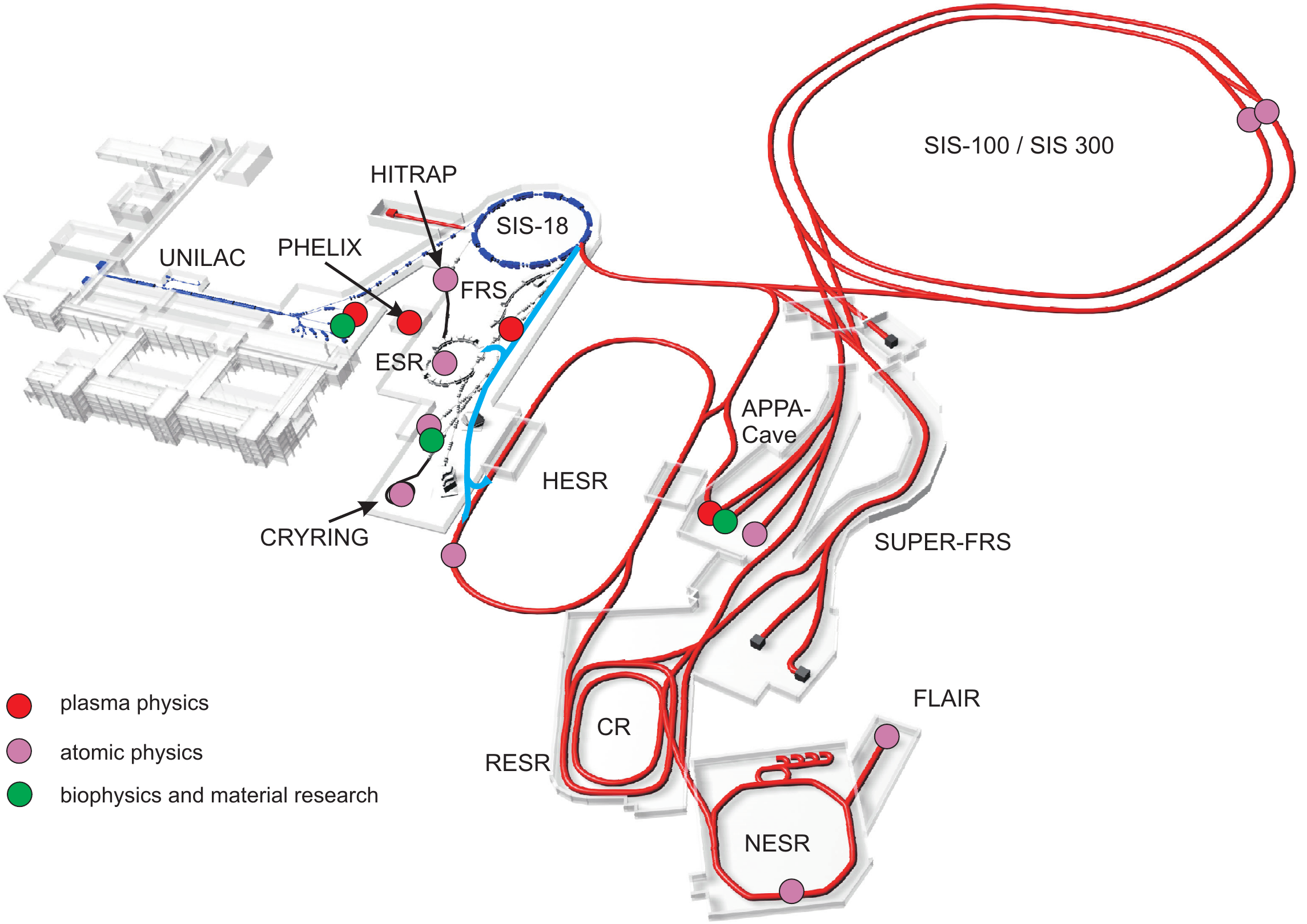}
\caption{A schematic view of the presently operational accelerator facility at GSI Helmholtz Center (gray) and the future FAIR facility (red).
The full version of FAIR is shown. The initial phase of FAIR, as defined by the MSV,
includes the SIS-100, APPA-Cave, CBM-Cave, Super-FRS, CR and HESR as well as connecting them beam lines.
The main locations of APPA experiments (plasma physics, atomic physics, biophysics and material research) are indicated in GSI and FAIR, including the HITRAP, which is being commissioned at the ESR, and CRYRING, which is presently under construction.
Possible  beam lines for transport of protons and ions from SIS-18 directly to the HESR and of antiprotons and ions from HESR to the ESR are shown with light-blue color. These beam lines are currently subject of detailed investigations.
\label{fig:1}}       % Give a unique label
\end{figure}

The FAIR project (see Figure~\ref{fig:1}) will be realized in stages as determined  by the Modularized Start Version (MSV) (see Ref. \cite{MSV}).
Since the New Experimental Storage Ring (NESR) \cite{NESR}, which is the main instrument for SPARC experiments in FAIR \cite{SPARC}, is not
within the first stage of the MSV, its realization will inevitably be delayed.
Therefore, the MSV has triggered substantial efforts to investigate alternatives,
enabling unique experiments in the realm of atomic physics using stored and cooled ion-beams already within the MSV.
In this paper we sketch the plans of the SPARC collaboration within the MSV as well as provide a glimpse beyond it.
Apart from the MSV program at a dedicated fix-target experimental hall, APPA-Cave, and laser-cooling experiments in SIS-100,
these plans include the installation of the CRYRING at the presently operating ESR \cite{CRYRING} and
the realization of an experimental program with relativistic ions beams in the High-Energy Storage Ring (HESR) \cite{HESR}.

\section{SPARC experiments at FAIR}
Realization of FAIR will allow for the extension of the atomic-physics research across virtually the full range of atomic matter,
i.e. concerning the accessible ionic charge states as well as beam energies \cite{ts_aip}.
SPARC experiments in two major research areas are planed:
collision dynamics in strong electro-magnetic fields and fundamental interactions between electrons and heavy nuclei up to bare uranium.

In the former area the relativistic heavy ions will be employed in a wide range of collision studies.
In the extremely short, relativistically enhanced field pulses, the critical field limit (Schwinger limit) for lepton-pair production can be surpassed
by orders of magnitudes and breakdowns of perturbative approximations for pair production are expected.
The detection methods of reaction microscopes will give the momentum of all fragments when atoms or molecules
are disintegrating in strong field pulses of the ions. This allows for exploring regimes of multi-photon processes that are still far from being reached with high-power lasers.
In particular, the storage ring HESR will be exploited for collision studies.
Here, fundamental atomic processes can be investigated for cooled heavy-ions at well-defined charge states interacting with photons, electrons and atoms.
These studies can even be extended at the CRYRING at ESR to the low-energy
regime where the atomic interactions are dominated by strong perturbations and quasi-molecular effects.

The other class of experiments will focus on structure studies of selected highly-charged ion species,
a field that is still largely unexplored; with determinations of properties of stable and unstable nuclei
by atomic physics techniques on the one hand, and precision tests of quantum electrodynamics (QED)
and fundamental interactions in extremely strong electromagnetic fields on the other hand.
Different complementary approaches such as relativistic Doppler boosts of optical or X-UV laser photons to the X-ray regime,
coherent radiation by channeling of relativistic ions, electron-ion recombination, and electron and photon spectroscopy will be used and will give hitherto unreachable accuracies. These transitions can also be used to laser-cool the relativistic heavy ions to extremely low temperature, which could lead to a break-through in accelerator technology.
Storing of unstable nuclides in a well-defined high atomic charge-state, that is in a well-defined leptons+nucleus quantum state,
will allow for high precision investigations of weak interaction through studying of $\beta$-decays under such clean conditions \cite{beta-decay,beta-decay2}.
Another important scenario for this class of experiments will be the slowing-down,
trapping and cooling of particles in the ion trap facility HITRAP \cite{HITRAP}.
This scenario will enable high-accuracy experiments in the realm of atomic and nuclear physics,
as well as highly-sensitive tests of the Standard Model.

For highly-charged heavy ions, FAIR will be worldwide unique with respect to the beam energies and intensities.
Fixed target experiments for highly-charged ions at relativistic energies with will be available only at FAIR.
In particular, the HESR will provide the possibility to exploit cooled relativistic ion beams.
Moreover, the use of storage rings for highly-charged ions is by itself a unique aspect.
Concerning the beam intensities, the storage rings encompass a large dynamical range providing highest intensities for cooled ion beams but accomplish also precision experiments with single stored ions.
At lower energies, comparable to the ESR, similar beam properties in terms of energies,
intensities and charge states are offered only at the heavy-ion accelerator and storage ring facility Lanzhou (China) \cite{Xiao}.
However, at Lanzhou no deceleration option for highly-charged ions as available for CRYRING and HITRAP exists at present.
One has to note, that a new-generation accelerator facility HIAF is planned in China,
where one of the research foci will be the physics with stored and cooled highly-charged ions at various energies \cite{HIAF}.
To some extend CRYRING might be compared with the Heidelberg TSR storage ring, which is planned to be installed at ISOLDE/CERN \cite{TSR}.
However, only at CRYRING the range of available charge states extends even to the heaviest bare ions, e.g. bare uranium.

\section{Recent Developments for Storage Ring Experiments at FAIR}
\label{sec:1}
\begin{figure}[t!]
\includegraphics[width=\textwidth]{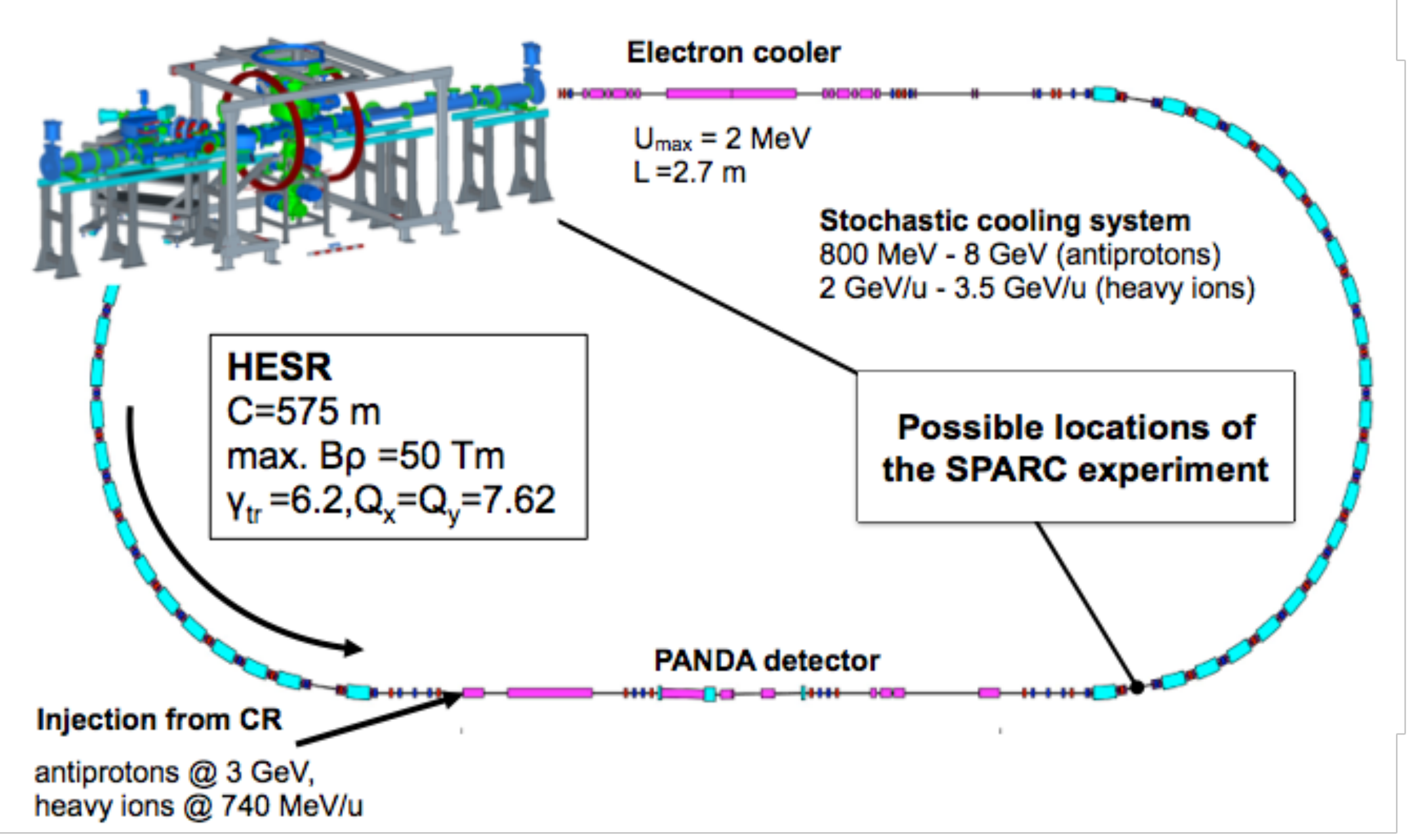}
\caption{The High-Energy Storage Ring HESR \cite{HESR}. The main parameters of the HESR are indicated as well as the locations of the PANDA and SPARC experiments. A scheme of the SPARC setup is shown the insert \cite{SPARCHESR,HCI}.
\label{fig:2}}       % Give a unique label
\end{figure}
The most recent technological developments at FAIR with respect to the atomic physics collaboration SPARC
are closely related to the novel physics opportunities.
These are provided by the heavy-ion storage-ring experiments at the HESR and by the reassembly of the CRYRING coupled to the current ESR. Whereas the high energy APPA cave offers direct SIS-100 beams of highly-charged heavy ions up to 10 GeV/u,
HESR can store cooled beams of highly-charged ions at energies of up to a few GeV/u.
This is an unique feature of FAIR, which is yet not feasible at any other place in the world.
The High-Energy Storage Ring (HESR) was primarily designed for experiments with stored and cooled anti-protons \cite{HESR}.
However, by detailed studies it turned out to be a well-suited facility which can accommodate a range of SPARC experiments with high-energy stored heavy-ion beams.
%The HESR can store cooled beams at energies of up to a few GeV/u and
%can thus enable unique atomic physics experiments which are not feasible at any other place in the world.
This is in particular true for the use of cooled ion beams at relativistic energies with $\gamma$-values ranging from 2 to 6,
an option unambiguously documented in a recent feasibility study \cite{SPARCHESR,HCI}.
The latter considered electron cooling, stochastic cooling, ion optical properties
at the foreseen location of the internal target as well as beam storage times relevant for the planned in-ring experiments (see Figure \ref{fig:2}).
Together with the specified, unrivalled properties of the HESR,
the frequencies of novel laser and laser-driven sources in the visible and the x-ray regime
can even be boosted in combined experiments with heavy ions.
Soft x-ray lasers, as developed for experiments at ESR and NESR,
will now give access to the study of transitions at much higher transition energies.
Especially, the interaction with highly-charged relativistic ions,
novel multi-keV photon sources will access new regimes of non-linear
photon matter interaction and the effects of QED in strong Coulomb fields. Even pump/probe laser experiments
for the investigations of decay properties of excited states of highly-ionized atoms can be anticipated.

Further physics topics to be addressed are: pair-production phenomena,
relativistic photon-matter interaction, correlated electron motion studied by target double-ionization,
tests of special relativity, bound state QED and nuclear parameters,
exotic nuclear decay modes in highly-charged ions, and parity non-conservation in high-Z ions and extreme electromagnetic fields.

As part of the Swedish in-kind contribution to FAIR,
CRYRING has long been foreseen for a relocation to Darmstadt, as integral instrument for the FLAIR facility \cite{FLAIR}.
In a Swedish-German joint effort, a project group has worked out a strategy for an early installation of CRYRING
in the evolving GSI/FAIR accelerator complex \cite{CRYRING}.
The study group report \cite{CRYRING} was evaluated and accepted as a Technical Design Report (TDR) by FAIR committees.
The transfer of CRYRING to GSI has already been completed and currently CRYRING is getting installed.
%A schematic view of the CRYRING is given in Figure~\ref{fig:3}.
Downstream of ESR an ideal location for setting up CRYRING was found (see Figure \ref{fig:1}),
where at minimum cost the access to beams of all ion species available at GSI
is guaranteed and opens an exciting opportunity for novel research.
The scientific program of CRYRING@ESR will focus on atomic and nuclear physics of exotic systems,
exploiting the available unique gas jet target and electron target, bridging the gap between the beam energies at the ESR ($< 4$~MeV/u) and at HITRAP ($< 5$~keV/q) \cite{CR_PB}.
The fast ramping capability of CRYRING will give access to intense beams of bare and exotic nuclei at low energies.
Thus, atomic collisions can be studied in the adiabatic regime by recoil ion momentum spectroscopy and,
in addition, the low Doppler shift/spread provides unique conditions for atomic structure investigations based on electron and photon detection.
With its independent RFQ injector beam line, CRYRING@ESR will serve as test bed for prototyping FAIR components in an operating environment.

%\begin{figure}
%\epsfig{file=fig3_cryring.eps,bbllx=120pt,bblly=355pt,bburx=700pt,bbury=720pt,height=8.cm,clip=}
%%  \includegraphics{example.eps}
%\caption{Schematic view of the CRYRING as it will be re-assembled at the ESR.}
%\label{fig:3}       % Give a unique label
%\end{figure}

\section{Further Activities of SPARC}

The R\&D activities of the SPARC collaboration at the new facility (within the MSV) are centered on the four main experimental areas for atomic physics: APPA-Cave, SIS-100, HESR as well as CRYRING and HITRAP coupled to the ESR.
The successful collaboration between various groups from different countries, which are working together
on the realization of the SPARC experimental program at FAIR by producing important new prototypes and advanced detectors,
is also emphasized by essential precursor experiments at the present experimental storage ring ESR of GSI and other suited, external facilities.
Such experiments allow for commissioning, testing and further improving of the equipment in real experiments.
For instance, decisive progress was achieved in ESR experiments employing powerful lasers:
\begin{itemize}
\item{A laser cooling experiment was conducted on 122 MeV/u C$^{3+}$ ions \cite{lasercooling}.
It demonstrated for the very first time cooling of a highly-charged ion down to
an unsurpassed longitudinal momentum spread  $\Delta p/p < 10^{-7}$ by exploration of the Doppler boost,
similar as it is finally planned for the experiments at HESR and SIS100/300.}
\item{Time Dilation in Special Relativity has been successfully tested using $^7$Li$^+$ at a velocity of $\beta =0.34$
\cite{time}.
The obtained results are in accordance with the most stringent previous tests.}
\item{Successful measurement of the hyperfine splitting in hydrogen- and lithium-like $^{209}$Bi through precision laser spectroscopy technique, allows on the one side for constraining the QED calculations and on the other side led to a development of efficient detector systems \cite{hyper}.}\\
\end{itemize}
Based on these and other experiments, the TDRs for laser experiments at SIS-100, HESR, and CRYRING are currently being worked on.
As further activities we like to mention:
\begin{itemize}
\item{A breakthrough was achieved by a very first high-resolution measurement of the Lyman-$\alpha$ transitions in high-Z
hydrogen-like systems as a precondition for a critical test of QED in the widely unexplored domain of very strong fields.
The measurements were carried out on hydrogen-like Au$^{78+}$ at the ESR storage ring and were made possible
by the development of the FOCAL x-ray crystal optics which overcomes both the limiting spectral resolving power.
The present efforts are an important pilot experiment for SPARC at FAIR emphasizing the physics of extreme electromagnetic fields \cite{FOCAL}.}
\item{The design, construction and installation of a cryogenic micro-calorimeter array ``maXs'',
dedicated for high resolution x-ray spectroscopy experiments at FAIR is progressing.
The maXs is based on the detection principle of metallic-magnetic calorimeters
and is operated at a temperature of about 30 mK. Besides its high-resolution properties it will also provide timing capability.
A TDR is ready for submission \cite{MAXs}. }
\item{Newly developed position-sensitive Ge(i) and Si(Li) detectors were taken into operation.
Detector performance tests were performed with photon beams at the ESRF, DORISIII, and PETRAIII synchrotron facilities.
A TDR is being worked on \cite{POLARI}. }
\item{Extensive investigations on the production of cryogenically cooled liquid hydrogen and helium droplet beams at the experimental storage ring ESR were carried out \cite{Nikos} with the goal to achieve high area densities for these low-Z internal targets. The results show that an area density of up to 10$^{15}$~cm$^{-2}$ is achieved for both light gases. Based on these systematic studies, a TDR for the internal target at HESR has been worked out and submitted \cite{GasJet}.}
\item{For the spectroscopy of electrons and positrons originating in collisions of very heavy, highly-charged ions with atoms/molecules/fiber targets in the relativistic domain, a concept for a reaction microscope in combination with a magnetic electron/positron spectrometer is currently being developed . A TDR will be prepared by the end of 2014.}
\item{For experimental studies concerning the electron dynamics in transiently formed superheavy quasi-molecules presently a combination of electron/recoil spectrometers is designed and built. This setup consists of a longitudinal reaction microscope for simultaneous position resolved spectroscopy of low energy electrons and recoil ions. The prototype of an imaging forward electron spectrometer has been installed and commissioned at the ESR \cite{electron}.}
\item{Newly developed non-destructive particle detectors employed for beam diagnostic purpose as well as for current calibration
purposes showed unparalleled sensitivity and very high time resolution \cite{Nolden11,Shahab}.
Furthermore, their sensitivity allows for deducing the longitudinal momentum kinematics in a reaction in a ring.
Together with Cryogenic-Current-Comparator-detectors \cite{CCC}, it will be possible to non-destructively monitor the beam currents for the entire dynamic range of intensities envisioned for SPARC Experiments in the HESR. The corresponding TDRs are being prepared.}
\item{As a feasibility test towards the future FAIR experiments, a new setup for high resolution x-ray spectroscopy of Li-like Uranium via resonant coherent excitation (RCE) has been installed in Cave-A and
has been successfully commissioned with a cooled 191.5 MeV/u U$^{89+}$ beam from the ESR to study the $1s^22s-1s^2p_{3/2}$ electron transition \cite{Nakano}.
A TDR for channeling experiments at SIS100 and HESR is under preparation.}
\item{HITRAP, a linear decelerator for heavy, highly-charged ions is installed downstream the ESR \cite{HITRAP}.
The deceleration is done in two steps, by using an interdigital H-type structure and a radio-frequency quadrupole.
The first part decelerates from 4 MeV/u to 500 keV/u and the second one further down to 6 keV/u.
Before being distributed to the experiments, the decelerated beam is captured in a Penning trap for cooling.
The facility is under commissioning and a number of online and offline tests have been recently performed.}
\end{itemize}

\section{Glimpse beyond the Modular Start Version}
%The low-energy antiproton physics community is making constant progress at the Antiproton Decelerator of CERN.
%Highlights in recent years were the first trapping of neutral antihydrogen atoms
%by the ALPHA collaboration and the first formation of antihydrogen in a cusp trap by the ASACUSA collaboration,
%which together were selected as the ``Highlight of the year 2010'' by Physics World.
%With ALPHA observing a first resonant hyperfine transition in trapped antihydrogen in 2011,
%ATRAP reporting also the trapping of antihydrogen in 2012, and ASACUSA preparing
%a polarised antihydrogen beam from the cusp trap, the phase of antihydrogen spectroscopy is about to start.
%Two experiments to measure the gravitation of antihydrogen, AEgIS and Gbar, were approved.
%ATRAP and a newly formed collaboration BASE are under way to measure the g-factor of the antiproton in a Penning trap.
%To cope with these increased activities, CERN has approved the construction of an additional storage ring
%called ELENA to more efficiently decelerate the antiproton beam from 5 MeV to 100 keV,
%and for opening the possibility of using an internal target for collision studies.
%A TDR is currently being prepared; the completion of ELENA is expected in 2017.

The current developments at Antiproton Decelerator of CERN demonstrate the rich science case for low-energy anti-proton physics as anticipated by the FLAIR collaboration at FAIR.
With CRYRING@ESR two fully commissioned storage rings will be available, and,
by installing an anti-proton transfer line, the physics program of the FLAIR collaboration could be realized at a very early stage. This option was recently (in spring 2012) discussed at a FLAIR workshop at GSI. One may note that a further important facility for FLAIR, HITRAP at the ESR, is currently getting commissioned.  This portfolio of facilities (CRYRING, HITRAP and the USR \cite{USR})
 will enable novel physics opportunities such as slowly extracted anti-proton beam which are even not covered by ELENA at AD.
 We would like to note, that there is a strong physics case pursued by the hadron-physics community to study doubly-antikaonic clusters \cite{Kienle},  which would profit from having slow antiproton beams at CRYRING.
 A possible beam line could connect the HESR with the ESR as, e.g., indicated in Figure \ref{fig:1}.
 In such a case the cooled and slowed-down antiprotons would be extracted from the HESR at 9.5 Tm towards the ESR, where they
 would further be cooled and slowed-down to about 1.4 Tm, the injection rigidity of the CRYRING, and transferred to CRYRING.

To facilitate the commissioning of the various machines of the FAIR facility, a direct beam line connecting SIS-18 and HESR could be imagined.
This would allow for an easier commissioning of the HESR on the one side.
On the other side, since the HESR is capable to efficiently accelerate the stored beams,
this would enable the exciting SPARC physics program at a very early stage of FAIR,
even before the commissioning of the complex accelerator chain SIS18-SIS100-CR-HESR is completed.
A possible location of such a beam line is indicated in Figure \ref{fig:1}.

Ultra-high power lasers combined with heavy-ion beams at FAIR represent a novel access to atomic physics and high-field physics.
Building on the well-established expertise within the Helmholtz centers and Helmholtz Institute Jena,
the Helmholtz Center Dresden/Rossendorf has proposed a ``Helmholtz-Beamline for FAIR'',
which is part of the Helm\-holtz-Road\-map for future research infrastructures (FIS, ``Forschungsinfrastrukturen'').
Coupled to various experimental stations of the FAIR accelerator complex,
this will lead to a significant expansion of the experimental options at FAIR for the realm of atomic and plasma physics.
A detailed feasibility study has to be worked out and submitted until 2016.

\begin{acknowledgements}
Helpful discussions and support by
Norbert Angert, Oleksii Dolinskii, Bernhard Franzke, Siegbert Hagmann, Takeshi Katayama, Oliver Kester, Edgar Mahner, Fritz Nolden, Klaus Peters, G{\"u}nther Rosner, Lars Schmitt, Boris Yu. Sharkov, \"Orjan Skeppstedt, Rolf Stassen, 
%Markus Steck, 
Horst St{\"o}cker, and Hans Stockhorst are gratefully acknowledged. This work was in part supported by the Helmholtz/CAS Joint Research Group HCJRG (Group No. HCJRG-108) and by the Helmholtz Alliance Program of the Helmholtz Association, Contract No. HA216/EMMI ``Extremes of Density and Temperature: Cosmic Matter in the Laboratory''.
\end{acknowledgements}

%\bibliographystyle{aps-nameyear}      % American Physical Society (APS) style, author-year citations
%\bibliography{example}                % name your BibTeX data base
%\nocite{*}


\end{document}